\documentstyle[12pt]{article}
\begin{document}
\newcommand{\be}{\begin{equation}}
\newcommand{\ee}{\end{equation}}
\newcommand{\ba}{\begin{eqnarray}}
\newcommand{\ea}{\end{eqnarray}}
\newcommand{\bi}{\bibitem}
\def\Res{\mathop{\rm Res}\nolimits}
\def\Re{\mathop{\rm Re}\nolimits}
\def\Im{\mathop{\rm Im}\nolimits}
\begin{titlepage}
\hbox to \hsize{\large \hfil  IHEP 99-23}
\hbox to \hsize{\hfil hep-ph/9906304}
\hbox to \hsize{\hfil April, 1999}
\vfill
\large \bf
\begin{center}
NONPERTURBATIVE POWER CORRECTIONS IN $\bar\alpha_s(q^2)$\\
OF TWO-LOOP ANALYTIZATION PROCEDURE 
\end{center}
\vskip 1cm
\normalsize
\begin{center}
{\bf Aleksey I. Alekseev
\footnote{Electronic address: 
alekseev@mx.ihep.su}}\\
{\small Institute for High Energy Physics,\\
142284 Protvino, Moscow Region, Russia}\\
\end{center}
\vskip 1.5cm
\begin{abstract}
The analytization procedure which allows one to remove
nonphysical singularities of the QCD running coupling constant
$\bar\alpha_s(q^2)$ in the infrared region is applied to standard
as well as to iterative solutions of the two-loop renormalization
group equation. Non-leading  at large  momentum 
nonperturbative contributions in $\bar\alpha_s(q^2)$ are obtained
in an explicit form. The coefficients of nonperturbative 
contributions
expansions in inverse powers of the squared Euclidean momentum are 
calculated.
For both cases considered there appear convergent at $q^2>\Lambda^2$
power series of negative terms with different dependence on term
numbers.
\end{abstract}
\vskip 1cm
PACS number(s): 12.38.Aw, 12.38.Lg
\vfill
\end{titlepage}
\section{Introduction}
The nonperturbative contributions to observables as well as 
direct to the QCD running coupling constant 
$\bar\alpha_s$ have been  widely discussed in recent years.
The evidence 
provided by very different approaches suggests the existence of
power type  nonperturbative corrections which reflect the complicated
structure of the QCD vacuum.

The nonperturbative contributions arise quite naturally when one
uses an analytization procedure. The main purpose of this procedure
is to remove nonphysical singularities from  approximate 
(perturbative) expressions for the Green functions of QFT. The idea
of the procedure goes back to Refs.~\cite{Red,Bog} devoted to the
ghost pole problem in QED. The foundation of the procedure is the
principle of summation of information derived from the perturbation 
theory  under the sign
of the K\"allen -- Lehmann spectral integral.
In recent papers~\cite{Shir,Shir1} it is suggested to solve the ghost
pole problem in QCD by demanding the 
$\bar\alpha_s(q^2)$ to be analytical in $q^2$ (to compare with 
dispersive approach~\cite{Doksh}). 
As a result, instead of the one-loop expression 
$\bar\alpha^{(1)}_s(q^2)=(4\pi/b_0)/\ln (q^2/\Lambda^2)$
which takes into account the leading logarithms and has the ghost
pole at $q^2=\Lambda^2$ ($q^2$ is the Euclidean momentum squared),
one obtains the  expression
\be
\bar\alpha^{(1)}_{an}(q^2)=\frac{4\pi}{b_0}\left[\frac{1}{\ln(q^2/
\Lambda^2)}+\frac{\Lambda^2}{\Lambda^2-q^2}\right].
\label{a1}
\ee
Eq.~(\ref{a1}) is an analytic function in the complex $q^2$-plane
with a cut along the negative real semiaxis. The pole of the
perturbative running coupling at $q^2=\Lambda^2$ is canceled by the
nonperturbative contribution 
$(\Lambda^2\mid_{g^2\rightarrow 0}$ $\simeq \mu^2\exp
\{-(4\pi)^2/(b_0g^2)\})$ and the value $\bar\alpha^{(1)}_{an}(0)=
4\pi/b_0$ appeared finite and independent of $\Lambda$. 
The nonperturbative contribution in Eq.~(\ref{a1}) can be presented 
as convergent at $q^2>\Lambda^2$ constant signs series in the inverse
powers of the momentum squared. 

At the two-loop level one cannot succeed in finding the 
nonperturbative part of
$\bar\alpha^{(2)}_{an}(q^2)$ in an explicit form. 
There were obtained the approximate formulas~\cite{Shir1,Shirkov}
which, in spite of their accuracy, do not provide an exhaustive
information on the behavior of the nonperturbative  contributions.
The most important feature of the analytization procedure 
discovered is the stability property of the value of the 
analytized coupling constant at zero with respect to high 
corrections, 
$\bar\alpha^{(1)}_{an}(0)=$	$\bar\alpha^{(2)}_{an}(0)=$
$\bar\alpha^{(3)}_{an}(0)$.
This property provides a definite high corrections stability of 
$\bar\alpha_{an}(q^2)$
in the infrared region in a hall.

In this paper we find out what happens with power corrections
of nonperturbative ultraviolet "tail" of the analytized running
coupling when going from the one-loop level to the two-loop level.

\section{The extraction of nonperturbative  contributions}
Let us consider the two-loop $\beta$-function
\be
\beta(g^2)=\beta_0g^4+\beta_1g^6,
\label{1}
\ee
where the coefficients
$$
\beta_0=-\frac{1}{(4\pi)^2}b_0, \,\,\, b_0=11-\frac{2}{3}n_f,
$$
\be
\beta_1=-\frac{1}{(4\pi)^4}b_1, \,\,\, b_1=102-\frac{38}{3}n_f
\label{2}
\ee
do not depend on the renormalization scheme choice. We can present
the Gell-Mann -- Low equation
\be
u\frac{\partial \bar g^2(u,g)}{\partial u}=\beta(\bar g^2),
\label{3}
\ee
$u=q^2/\mu^2$ in a "rationalized" form by introducing the function
$a(x)$ of $x=q^2/\Lambda^2$ of the form
$$
a(x)=
\frac{b_0}{(4\pi)^2}
\bar g^2(u,g)=
\frac{b_0}{4\pi}\bar\alpha_s(q^2).
$$
The two-loop differential equation for the running coupling constant
can be written as follows:
\be
x\frac{da(x)}{dx}=-a^2(x)-ba^3(x),
\label{5}
\ee
where
\be
b=-\frac{\beta_1}{\beta^2_0}=\frac{102-\frac{38}{3}n_f}
{(11-\frac{2}{3}n_f)^2}.
\label{4}
\ee
For the case of three active quark flavors $n_f=3$, $b_0=9$, and
$b=64/81\simeq 0.7901$.
Integrating Eq.~(\ref{5}) one obtains
\be
\frac{1}{a(x)}-b\ln \left(1+\frac{1}{ba(x)}\right)=\ln x+\ln c,
\label{6}
\ee
where $c$ being the integration constant.
The transcendental Eq.~(\ref{6})  defines implicitly the function
$a(x)$. The real solution of Eq.~(\ref{6}) has two branches,
the former  corresponds to  a positive decreasing at
$x\rightarrow\infty$ function $a(x)$, whereas for the latter
$a(x)$ is negative and at
$x\rightarrow\infty$ it goes to  $-1/b$. 
The positive decreasing at $x\rightarrow\infty$ solution corresponds 
to the asymptotic  freedom property, and this solution can be fixed
further by the choice of the integration constant $c$, which means,
in practice, the  redefinition of $\Lambda$.
The exact solution of Eq.~(\ref{6})  can be written~\cite{Magradze} 
using the Lampert function $W(y)$ defined by the equation
$y=W(y)\exp \{W(y)\}$.

We shall consider the following two forms of approximate solutions
of Eq.~(\ref{6}). The first one is the standard two-loop coupling
\be
a^{(2)}(x)=\frac{1}{l}-\frac{b}{l^2}\ln l,
\label{9}
\ee
where $l=\ln x$, and the second one is the iterative 
solution ~\cite{Shirkov} of the form
\be
a^{it}(x)=\frac{1}{l+b\ln \left(1+\frac{l}{b}\right)}.
\label{8}
\ee
This solution, when expanding in inverse powers of logarithms
apart from the terms Eq.~(\ref{9}), gives rise to the term
$b\ln b/l^2$ and additional terms of the form
$l^{-3}\ln^2l+...$ which should be taken into account at the
three-loop level. The exact solution of Eq.~(\ref{6}) also does not
claim to be  the description of the three-loop terms. At large 
$x$ the functions~(\ref{9}), (\ref{8}) behave similarly but 
for small $x$ the behavior is different and at
$x=1$ they have singularities  of a different analytical structure.
Namely, at $x\simeq 1$ 
\be
a^{(2)}(x)\simeq-\frac{b}{(x-1)^2}\ln (x-1),\,\,\,
a^{it}(x)\simeq\frac{1}{2(x-1)}.
\label{10}
\ee
These singularities are stronger than the integrable singularity
of the exact solution of Eq.~(\ref{6}) of the form 
$1/\sqrt{x-1/c}$. But this is not an obstacle to the application 
of the analytization procedure.

According to the definition the analytized running coupling is 
obtained from the initial running coupling by the integral
representation
\be
a_{an}(x)=\frac{1}{\pi}\int\limits_0^\infty \frac{d\sigma}{x+\sigma}
\rho(\sigma),
\label{15}
\ee
where $\rho(\sigma)=\Im a_{an}(-\sigma-i0)=\Im a(-\sigma-i0)$.
As  initial expressions we will consider Eqs.~(\ref{9}), (\ref{8}).
By making the analytic continuation of these equations into 
the Minkowski
space $x=-\sigma-i0$, one  obtains,	 correspondingly
\be
a^{(2)}(-\sigma-i0)=\frac{1}{\ln \sigma-i\pi}-\frac{b}{(\ln \sigma
-i\pi)^2}\ln\left(\ln\sigma-i\pi\right),
\label{11}
\ee
$$
a^{it}(-\sigma-i0)=\frac{1}{\ln\sigma-i\pi+b\ln\left(1+\frac{1}{b}
\ln\sigma-\frac{i\pi}{b}\right)}=
$$
\be
=\frac{1}{\ln \sigma+b\ln \sqrt{\left(1+\frac{1}{b}\ln\sigma\right)^2
+\frac{\pi^2}{b^2}}-i\left[\pi+b\arctan \frac{\pi}{b+\ln \sigma}
\right]}.
\label{12}
\ee
For the spectral function $\rho(\sigma)$ it follows that
$$
\rho^{(2)}(\sigma)=\frac{\pi}{\ln^2\sigma+\pi^2}+\frac{b}{\left(
\ln^2\sigma +\pi^2\right)^2}\times
$$
\be
\times\left[\left(\ln^2\sigma-\pi^2\right)\arctan \frac{\pi}{\ln
\sigma}-2\pi\ln\sigma\ln \sqrt{\ln^2\sigma+\pi^2}\right],
\label{13}
\ee
\be
\rho^{it}(\sigma)=\frac{\pi+b\arctan\frac{\pi}{b+\ln\sigma}}{
\left[\ln\sigma+b\ln\sqrt{\left(1+\frac{1}{b}\ln\sigma\right)^2+
\frac{\pi^2}{b^2}}\right]^2 + \left[\pi+b\arctan\frac{\pi}{b+
\ln\sigma}\right]^2}.
\label{14}
\ee
By the change of the variable of the form $\sigma=\exp (t)$,
the analytized expressions
are derived from~(\ref{13}), (\ref{14}) as follows:
$$
a^{(2)}_{an}(x)=\frac{1}{\pi}\int\limits^\infty_{-
\infty}dt\,\frac{e^t}{
x+e^t}\times
$$
\be
\times\left\{\frac{\pi}{t^2+\pi^2}+\frac{b}{
\left(t^2+\pi^2\right)^2}
\left[\left(t^2-\pi^2\right)\arctan\frac{\pi}{t}-
2\pi t\ln\sqrt{t^2+\pi^2}\right]\right\},
\label{16}
\ee
\be
a^{it}_{an}(x)=\frac{1}{\pi}\int\limits^\infty_{-\infty}
dt\,\frac{e^t}{
x+e^t}
\frac{\pi+b\arctan\frac{\pi}{b+t}}{\left[t+b\ln\sqrt{\left(
1+\frac{t}{b}\right)^2+\frac{\pi^2}{b^2}}\right]^2
+\left[\pi+b\arctan\frac{\pi}{b+t}\right]^2}.
\label{17}
\ee
Eqs.~(\ref{16}), (\ref{17}) can be used to study 
$a_{an}(x)$ by numerical methods in particular. We are interested in
nonperturbative contributions which decrease  at
$x\rightarrow\infty$ much  faster than the perturbative ones,
therefore, their extraction by a direct comparison with the initial
expressions is difficult at large $x$. We shall obtain an explicit
analytic formulae for the nonperturbative contributions 
as a difference
between output and input expressions of the analytization procedure.

Let us see what the singularities of the integrands of~(\ref{16}), 
(\ref{17}) in the complex  $t$-plane are.
First of all, for both cases the integrands have the simple
poles at  $t=\ln x\pm i\pi(1+2n)$,
$n=0,1,2,...$. All the residues of the function  
$\exp(t)/(x+\exp(t))$ 
at these points are equal to unity. Making use of the formula 
\be
\arctan\frac{\pi}{t}=\frac{1}{2i}\ln\frac{t+i\pi}{t-i\pi}
\label{18}
\ee
one can obtain 
$$
\frac{1}{\left( t^2+\pi^2\right)^2}\left[\left(t^2-\pi^2\right)
\arctan\frac{\pi}{t}-2\pi t\ln \sqrt{t^2+\pi^2}\right]=
$$
\be
=\frac{i}{2}\left[\frac{\ln(t-i\pi)}{(t-i\pi)^2}-
\frac{\ln(t+i\pi)}{(t+i\pi)^2}\right].
\label{19}
\ee
This formula allows one to make transparent the singularities 
structure of the integrand in~(\ref{16}). We obtain
$$
a^{(2)}_{an}(x)=\frac{1}{2\pi i}\int\limits^\infty_{-\infty}
dt \, \frac{e^t}{x+e^t}\times
$$
\be
\times\left\{\frac{1}{t-i\pi}-\frac{1}{t+i\pi}-b
\left[\frac{\ln(t-i\pi)}
{(t-i\pi)^2}-\frac{\ln(t+i\pi)}{(t+i\pi)^2}\right]\right\}.
\label{20}
\ee
For the second  case under  consideration, it follows 
from~(\ref{12}),
(\ref{17}) that
$$
a^{it}_{an}(x)=\frac{1}{2\pi i}\int\limits^\infty_{-\infty}
dt \, \frac{e^t}{x+e^t}\times
$$
\be
\times \left[\frac{1}{t-i\pi+b\ln\left(1+\frac{t}{b}-\frac{i\pi}
{b}\right)}-\frac{1}{t+i\pi+b\ln\left(1+\frac{t}{b}+
\frac{i\pi}{b}
\right)}\right].
\label{21}
\ee
The integrands in~(\ref{20}), (\ref{21}) multiplied by $t$ go to
zero at $\mid t\mid\rightarrow \infty$. This allows one to append
the integration by the arch of the "infinite" radius without 
affecting the value of the integral. Closing the integration contour
$C$ in the upper half-plane of the complex variable $t$,
by using the residue theorem for the one-loop part of~(\ref{20}), 
one readily obtains  the well-known result~(\ref{a1})
$$
a^{(1)}_{an}(x)=\frac{1}{2\pi i}\int\limits_C dt\,f(t)=
\sum\limits^\infty_{n=0}\Res f\left(t=\ln x+i\pi(1+2n)\right)+
$$
\be
+\Res f\left(t=i\pi\right)=\frac{1}{\ln x}+\frac{1}{1-x},
\label{22}
\ee
where  $f(t)$ is the integrand of~(\ref{20}) at $b=0$.
The two-loop part of the integrand~(\ref{20}) apart from the simple
poles at  $t=\ln x \pm i\pi(1+2n)$, $n=0,1,2,...$
has logarithmic branch points at $t=\pm i\pi$ which coincide with
the second order poles. Let us cut the complex $t$-plane in a  
standard way, $t=\pm i\pi-\lambda$, with $\lambda$ being the real
parameter varying from  $0$ to $\infty$.
For the case of coincidence	of the branch point and the pole of the
second order the following theorem can be proved.

Theorem. \it
If  $f(z)$ analytic function inside the circle of the radius $r$
with the center at zero, then
\be
\int\limits_C dz\,\frac{\ln z}{z^2}f(z)=-2\pi i\left\{\int\limits
^0_{-a}\frac{dx}{x^2}\left[f(x)-f(0)-xf'(0)\right]-\frac{1}{a}f(0)-
f'(0)\ln a\right\},
\label{23}
\ee
where the contour $C$ goes from  $z=-a-i0$ along the lower side 
of the cut till the circle of the radius  $\delta$ with the center
at zero ($r>a>\delta>0$), then goes by this circle around  $z=0$, 
and then along the upper side of the cut it goes to  $z=-a+i0$. 
\rm

For the terms of~(\ref{20}) which are proportional to $b$,
we also close the contour in the upper half-plane of the complex 
variable $t$ excepting the singularity at $t=i\pi$. In this case 
an additional contribution emerges because of the integration along
the sides of the cut and around the second order pole. The
substitution  $z=t-i\pi$ brings this contribution to the form 
corresponding to the theorem~(\ref{23}) with  $f(z)=(b/2\pi i)
\exp(z)/(x-\exp(z))$. 		We shall call
this contribution together with the 
one-loop contribution from the pole at $t=i\pi$ as
nonperturbative because  the remaining contribution of the poles
at  $t=\ln x+i\pi(1+2n)$, $n=0,1,2,...$ results exactly in 
perturbative expression~(\ref{9}). In fact, according to the
residue theorem 
$$
a^{(2)\,pt}_{an}(x)=a^{(1)\,pt}_{an}(x)-b\sum\limits^\infty_
{n=0}\left[\frac{\ln (\ln x+2\pi i n)}{\left(\ln x+2\pi in\right)^2}
- \frac{\ln (\ln x+2\pi i (n+1)}{\left(\ln x+2\pi i(n+1)\right)^2}
\right]=
$$
\be
=\frac{1}{\ln x}-b\frac{\ln(\ln	x)}{(\ln x)^2}.
\label{24}
\ee
The nonperturbative part of the analytized coupling using 
Eqs.~(\ref{20}), (\ref{23}), (\ref{22}) can be derived with the
result
$$
a^{(2)\,npt}_{an}(x)=-\frac{1}{x-1}-b\left\{\int\limits^1_0
\frac{dt}{t^2}\left[\frac{1}{xe^t-1}-\frac{1}{x-1}+\frac{tx}{(x-1)^2}
\right]\right.+
$$
\be
+\left.\int\limits^\infty_1\frac{dt}{t^2}\frac{1}{xe^t-1}-\frac{1}
{x-1}\right\}.
\label{25}
\ee
Integrating by parts and making the change of variable 
$\sigma=\exp(-t)$, we can represent~(\ref{25}) in the form of
one integral with finite limits
\be
a^{(2)\, npt}_{an}(x)=\frac{1}{1-x}+b\left\{\frac{x}{(1-x)^2}+
x\int\limits^1_0d\sigma\, \ln(-\ln\sigma)\frac{x+\sigma}{(x-
\sigma)^3}\right\}.
\label{26}
\ee
This formula is convenient for the numerical study whereas 
formula~(\ref{25}) is suitable for finding the explicit power
nonperturbative corrections at $x\rightarrow\infty$.

Let us now turn to the evaluation of the integral~(\ref{21}) for
the analytized iterative coupling. The integrand has the simple
poles at  $t=\ln x\pm i\pi(1+2n)$, $n=0,1,2,...$ and at $t=\pm i\pi$.
It has also the logarithmic branch points at $t=-b\pm i\pi$.
To provide the single-valuedness of the integrand, we draw two cuts
beginning at  $t=-b\pm i\pi$ and going to  infinity to the left in
parallel to the real axis of the complex $t$-plane. We close the 
integration contour by the "infinite" semicircle in the upper 
half-plane with going around the branch point by the sides of the cut
and integrate along this contour. As in the previous case, all the 
terms in the sum of residues at  $t=\ln x+\pi i(1+2n)$, 
$n=0,1,2,...$ mutually cancel apart from the term of $n=0$.
Integrating also along the sides of the cut, we obtain 
$$
a^{it}_{an}(x)=\frac{1}{\ln x+b\ln \left(1+\frac{1}{b}\ln x\right)
}+\frac{1}{2(1-x)}+
$$
\be
+\int\limits^\infty_0 d\xi\, \frac{1}{1-xe^{b(1+\xi)}}
\frac{1}{\left(1+\xi-\ln\xi\right)^2+\pi^2}.
\label{27}
\ee
Performing the change of variable $\sigma=\exp(-\xi)$, we come 
to the integral with finite limits. For additional "nonperturbative"
output of the analytization procedure we obtain the following
representation
\be
\Delta a^{it}_{an}(x)=\frac{1}{2(1-x)}+\int\limits^1_0\frac{d\sigma}{
\sigma}\frac{1}{1-x(\sigma/e)^{-b}}\frac{1}{\left[1-\ln(-\sigma\ln 
\sigma)\right]^2+\pi^2}.
\label{28}
\ee

\section{Nonperturbative contributions at large $q^2$.}
Consider the large  $x$ behavior of the nonperturbative 
contributions. It is seen from Eq.~(\ref{25}) that this behavior
is regular. Expanding the integrands in powers of $1/x$ and
integrating, we find
\be
a^{(2)\, npt}_{an}(x)=\sum\limits^\infty_{n=1}\frac{c_{n}}
{x^n},
\label{29}
\ee
where
\be
c_{n}=-1+bn\left[n\int\limits^\infty_0dt\,\ln t e^{-nt}+1
\right].
\label{30}
\ee
Performing the $t$ integration in~(\ref{30}) yields the following
simple formula for the coefficients of the ultraviolet expansion
of the nonperturbative contributions
\be
c_{n}=-1+bn\left(1-\gamma-\ln n\right),
\label{32}
\ee
where  $\gamma$ is the Euler constant,  $\gamma\simeq 0.5772$.
The leading term is 
\be
c_{1}=-1+b(1-\gamma),
\label{320}
\ee
and the passing from the one-loop level to the two-loop one
results in some compensation of the leading at large $x$ term of the
form $1/x$. For the next terms there is no compensation, instead
they increase with $n$ increasing. Estimating at
$n_f=3$ the first three terms of the series~(\ref{29}), we obtain
\be
a^{(2)\,npt}_{an}(x)\simeq -\frac{0.666}{x}-\frac{1.427}{x^2}-
\frac{2.602}{x^3}- ...\,.
\label{33}
\ee

Let us consider the large $x$ behavior of the "nonperturbative"
contribution in~(\ref{27}). Expanding the integrand in powers of
$1/x$, we also obtain the infinite terms series 
\be
\Delta a^{it}_{an}(x)=\sum\limits^\infty_{n=1}\frac{c^{it}_{n}}
{x^n},
\label{34}
\ee
where
\be
c^{it}_{n}=-\frac{1}{2}-\int\limits^\infty_0d\xi \,e^{-bn(1+\xi)}
\frac{1}{\left[1+\xi-\ln \xi\right]^2+\pi^2}.
\label{35}
\ee
Expression~(\ref{35}) for the expansion coefficients
is not so simple as~(\ref{32}) but it is explicit and can be 
calculated up to an arbitrary accuracy. We see from Eq.~(\ref{35})
that all  $c^{it}_{n}$ are also negative tending fast to $-1/2$
when $n$ increases. Substituting the denominator in the integrand 
of~(\ref{35}) by its minimal value  $\pi^2+4$, one finds
\be
\mid c^{it}_{n}+
\frac{1}{2}\mid<\frac{1}{\pi^2+4}\frac{e^{-bn}}{bn}.
\label{36}
\ee
At $n_f=3$ it gives for the first coefficient $\mid c^{it}_{1}+1/2
\mid<0.0414$.
For the first three terms of the series~(\ref{34}), one can obtain
\be
\Delta a^{it}_{an}(x)\simeq -\frac{0.535}{x}-\frac{0.508}{x^2}-
\frac{0.502}{x^3}- ...\,.
\label{37}
\ee
Thus, passing from the one-loop level of the analytization 
procedure to the two-loop level leads to approximately two times
decrease  of each term of the expansion~(\ref{34}).

\section{Conclusion}
For both cases considered the nonperturbative contributions can be
presented in the form of series in inverse powers of the momentum
squared, all the terms are negative. Although the convergence radii 
of the series~(\ref{29}), (\ref{34}) are the same and equal to unity,
the dependence of the expansion coefficients on term numbers is
different. This is defined by the expressions taken as an input for
the analytization procedure. Choosing as an input the standard 
two-loop expression~(\ref{9}), we obtain the "standard 
nonperturbative corrections" of the form~(\ref{33}). 
For sufficiently large $x=q^2/\Lambda^2$ the nonperturbative 
contributions are defined mainly by the leading terms in~(\ref{33}),
(\ref{37}). The partial compensation of the leading terms found for
both cases considered when we pass from the one-loop level to the
two-loop level can point to the tendency of a high-loop minimization
of the nonperturbative contributions in the ultraviolet region.

The analytization procedure  may be not an
ultimate step  of construction the "physical" running coupling
constant. In terms of Ref.~\cite{Grun} the analytized coupling
constant corresponds to the "regularized perturbative part"
of the full coupling constant which contains also the "genuine
nonperturbative" contribution.
As a step in the direction of the full running coupling constant
it has been suggested in~\cite{AlekArbYF98,AlekArbMPL98} that
the one-loop analytization output be modified in a minimal way 
by introducing
the two additional pole type nonperturbative terms to provide
the ultraviolet convergence of the gluon condensate. In this case
the model  for the running coupling  constant arises with the
enhancement at zero momentum and dynamical gluon mass $m_g$.
This mass can be fixed~\cite{Australia}  by the condition of 
minimum of the nonperturbative vacuum energy and is estimated
$m_g\simeq $ 0.6 GeV for the "standard" value of the gluon
condensate (0.33~GeV)$^4$. The next step of the model building
for the full running coupling constant on the base of the
two-loop analytized  running coupling is under  study.

I am indebted to B.A.~Arbuzov, V.A.~Petrov, V.E.~Rochev for 
useful discussions. This work was supported in part by RFBR
under Grant No.~99-01-00091.

\end{document}